\documentclass[11pt]{article}
\usepackage{moriond,epsfig}

\def\be{\begin{equation}}
\def\ee{\end{equation}}
\def\bea{\begin{eqnarray}}
\def\eea{\end{eqnarray}}

\begin{document}
\title{F/S INTERFACES: POINT CONTACT VERSUS\\ ATOMIC THICKNESS GEOMETRIES}

\author{\underline{R. M\'ELIN}}

\address{Centre de Recherches sur les Tr\`es Basses
Temp\'eratures (CRTBT~\footnote{U.P.R. 5001 du CNRS, Laboratoire
conventionn\'e avec l'Universit\'e Joseph Fourier}),\\ CNRS, BP 166,
38042 Grenoble Cedex 9, France}

\maketitle
\abstracts{
We contrast perturbative expansions of ferromagnet / superconductor interfaces
in two geometries: (i) a point contact geometry where a single weak link connects
a 3D ferromagnet to a 3D superconductor and (ii) an atomic thickness geometry
with an infinite planar interface connecting a quasi-2D ferromagnet to a quasi-2D
superconductor. Perturbation theories are rather different in the two approaches
but they both break down at order $t^4$ ($t$ is the tunnel amplitude).
The regimes of strong ferromagnets are in
a qualitative agreement in both geometries. The regime of weak ferromagnets
exists only for the atomic thickness geometry and is related to Andreev bound states
due to lateral confinement in the superconductor.
}

\section{Introduction}

Many recent works have been devoted to
equilibrium properties of ferromagnet / superconductor (F/S) interfaces, which consists in
determining the value of the self-consistent superconducting gap as a function of the
various parameters. For instance it was shown that the critical temperature of F/S
superlattices is reentrant as a function of the exchange field~\cite{Andreev,Houzet}
and that the critical temperature of F/S bilayers is reentrant as a function
of the thickness of the ferromagnet~\cite{Baladie,Fominov,Proshin}. Other predictions were
made recently within a model where the superconductor is connected by a single
weak link to a ferromagnet~\cite{Melin-tr}: it was shown within this model that the superconducting
gap of a F/S/F trilayer in the parallel alignment is larger than in the
antiparallel alignment~\cite{Melin-FSF}, a result
that was also obtained within a model of F/S/F trilayer with atomic thickness
and half-metal ferromagnets~\cite{Buzdin-Daumens} that was finally extended to
Stoner ferromagnets~\cite{Melin-Feinberg}. The goal of this note is to compare
the perturbative expansions in the two approaches (point contact versus atomic
thickness geometries).

\section{Point contacts}

\subsection{The model}
\begin{figure}
\begin{center}
\psfig{figure=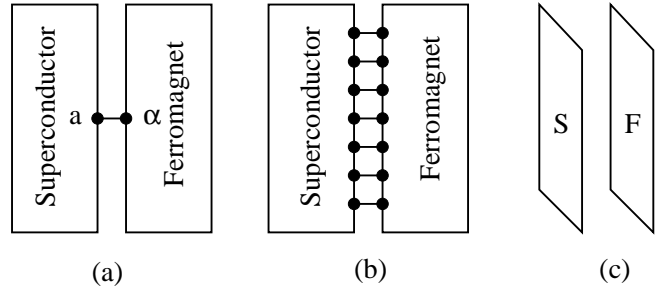,height=1.5in}
\end{center}
\caption{(a) Schematic representation of
different types of ferromagnet / superconductor contacts.
In (a) a 3D ferromagnet and a 3D superconductor are connected
by a single weak link~\protect\cite{Melin-FSF}.
In (b) a 3D ferromagnet and a 3D superconductor are connected
by an infinite planar contact~\protect\cite{Melin-FSF}.
In (c) an atomic thickness 2D ferromagnet is connected to an
atomic thickness 2D superconductor
by an infinite planar contact~\protect\cite{Buzdin-Daumens,Melin-Feinberg,Melin}.
\label{fig:schema}}
\end{figure}

Let us consider a three dimensional (3D) ferromagnet 
described by the Stoner model
connected by a single link
to a 3D superconductor described by the BCS Hamiltonian
(see Fig.~\ref{fig:schema}-(a)).
The Hamiltonian takes the form
${\cal H} = {\cal H}_{\rm BCS} + {\cal H}_{\rm Stoner} + {\cal W}$,
where the BCS Hamiltonian is given by
${\cal H}_{\rm BCS} = \sum_{\bf k} \epsilon_k c_{{\bf k},\sigma}^+
c_{{\bf k},\sigma} + \Delta \sum_{\bf k} \left(
c_{ {\bf k},\uparrow}^+ c_{ -{\bf k},\downarrow}^+
+c_{ {\bf k},\downarrow} c_{-{\bf k},\uparrow} \right)$,
the Stoner Hamiltonian is given by
${\cal H}_{\rm Stoner} = \sum_{ {\bf k},\sigma}
\epsilon_k c_{{\bf k},\sigma}^+ c_{{\bf k},\sigma}
-h_{\rm ex} \sum_{{\bf k},\sigma} \left(
c_{{\bf k},\uparrow}^+ c_{{\bf k},\uparrow} - 
c_{{\bf k},\downarrow}^+ c_{{\bf k},\downarrow}\right)$
and the tunnel Hamiltonian is given by
${\cal W} = t \sum_\sigma \left( c_\alpha^+ c_a+
c_a^+ c_\alpha \right)$ where $\alpha$ and $a$ are neighboring
sites belonging to the superconductor and ferromagnet respectively.

\subsection{Green's functions}

For the point contact model we use the Green's functions 
in real space:
\begin{equation}
\hat{g}(R,\omega) = \frac{2 m}{\hbar^2} \frac{1}{2\pi R}
\exp{\left(-\frac{R}{\xi(\omega)}\right)}
\left\{
\frac{\sin{(k_F R)}}{\sqrt{\Delta^2-\omega^2}}
\left[ \begin{array}{cc} -\omega & \Delta \\
\Delta & -\omega \end{array} \right]
+\cos{(k_F R)} \left[ \begin{array}{cc} -1 & 0 \\ 0 & 1 
\end{array} \right] \right\}
,
\end{equation}
where $\xi(\omega)=\hbar v_F/\sqrt{|\Delta^2-\omega^2|}$ is the
coherence length at a finite frequency, $\Delta$ is the superconducting
gap and $k_F$ is the Fermi wave-vector.
The local propagator $\hat{g}_{\rm loc}(\omega)$ is regularized by
introducing an ultra-violet cut-off $R_0$:
\begin{equation}
\hat{g}_{\rm loc}(\omega) = \frac{m}{\pi \hbar^2}
\left[ \begin{array}{cc} -k_F \omega/\sqrt{\Delta^2-\omega^2} -1/R_0
& k_F \Delta/\sqrt{\Delta^2-\omega^2} \\
k_F \Delta/\sqrt{\Delta^2-\omega^2} &
-k_F \omega/\sqrt{\Delta^2-\omega^2} +1/R_0 \end{array} \right]
.
\end{equation}

\subsection{Expansion of the magnetization in the superconductor}

The fully dressed Green's function $\hat{G}_{x,x}$ in the superconductor is
obtained through the Dyson equation $\hat{G}_{x,x}=\hat{g}_{x,x}
+\hat{g}_{x,\alpha} \hat{t}_{\alpha,a} \hat{g}_{a,a} \hat{t}_{a,\alpha}
\hat{G}_{\alpha,x}$, where $\hat{G}_{\alpha,x}$ is determined through
$\hat{G}_{\alpha,x}=\hat{g}_{\alpha,x} + \hat{g}_{\alpha,\alpha}
\hat{t}_{\alpha,a} \hat{g}_{a,a} \hat{t}_{a,\alpha} \hat{G}_{\alpha,x}$.
The magnetization $m(\omega)$ induced in the superconductor
is defined as the difference between the
spin-up and spin-down density of states. We find the expansion
$m(\omega)/h_{\rm ex} = A^{(2)}(\omega) t^2 + A^{(4)}(\omega) t^4 + ...$, with
\begin{eqnarray}
A^{(2)}(\omega) &=& \frac{m^4}{\pi^4 \hbar^6 k_F R^2}
\exp{\left(-\frac{2 R}{\xi(\omega)}\right)}
\left\{ \cos{(2 k_F R)} + \frac{\omega}{\sqrt{\Delta^2-\omega^2}} \sin{(2 k_F R)} \right\}\\
A^{(4)}(\omega) &=&  \frac{m^6}{\pi^6 \hbar^{10} R_0 R^2}
\exp{\left(-\frac{2 R}{\xi(\omega)}\right)}
\left\{ \frac{\omega}{\sqrt{\Delta^2-\omega^2}}
\cos{(2 k_F R)} - \sin{(2 k_F R)} \right\}\\\label{eq:A4-inf}
&+&  \frac{m^6}{\pi^6 \hbar^{10} k_F R_0^2 R^2}
\exp{\left(-\frac{2 R}{\xi(\omega)}\right)}
\left\{ \cos{(2 k_F R)} + \frac{\omega}{\sqrt{\Delta^2-\omega^2}} \sin{(2 k_F R)}   \right\}
.
\end{eqnarray}
The second order term shows $2 k_F$ Friedel oscillations and does not depend on the
ultra-violet cut-off $R_0$. The fourth order terms are divergent if
the ultra-violet cut-off $R_0$ tends to zero.

We find a similar expansion for $\omega>\Delta$. The structure of the expansion
is given by
\begin{eqnarray}
&& A^{(2)}(\omega) =\frac{m^4}{\pi^4 \hbar^6 k_F R^2}
\left\{ \cos{(2 k_F R)} \cos{\left(\frac{2 R}{\xi(\omega)}\right)}-
\frac{\omega}{\sqrt{\omega^2-\Delta^2}}
\sin{(2 k_F R)}
\sin{\left(\frac{2 R}{\xi(\omega)} \right)}
\right\}\\\label{eq:A4-sup}
&&A^{(4)}(\omega) = \frac{m^6}{\pi^6 \hbar^{10}}
\left\{ \frac{k_F}{R^2} A^{(4)}_0 + \frac{1}{R_0 R^2} A^{(4)}_1
+\frac{1}{k_F R_0^2 R^2} A^{(4)}_2 \right\}
,
\end{eqnarray} 
where we obtained the explicit expression of 
$A^{(4)}_0$, $A^{(4)}_1$ and $A^{(4)}_2$. The perturbative expansion
breaks down since the final result is diverging in the limit
$R_0 \rightarrow 0$.
The divergences can be removed empirically by interpreting $R_0$
as an extra integration variable in the fourth order diagrams.
The diverging prefactors cancel once the integration over $R_0$ is carried out.
For example $A^{(4)}(\omega)$ in (\ref{eq:A4-inf}) and (\ref{eq:A4-sup})
is replaced by $\int_0^{1/k_F} 4 \pi R_0^2 A^{(4)}(\omega) d R_0$.
However the proportionality factor in the
upper bound of the integral over $R_0$ remains arbitrary.

\subsection{Expansion of the Gorkov function}
The superconducting gap is
determined self-consistently through the relation
\begin{equation}
\label{eq:self}
\Delta_{\bf x} = U \int_0^D \frac{d \omega}{2 \pi}
\mbox{Im} \left[G^{1,2,A}_{{\bf x},{\bf x}}(\omega) \right]
,
\end{equation}
where $U$ is the microscopic attractive interaction, $D$ is the band-width
and $G^{1,2,A}_{{\bf x},{\bf x}}(\omega)$ is the advanced ``12'' component of the
Green's function.
The expansion of the Gorkov function is given by
$G_{1,2}(\omega) = g_{1,2}(\omega) + B^{(2)}(\omega) t^2 + B^{(4)}(\omega) t^4 +...$, with
\begin{eqnarray}
g_{1,2}(\omega) &=& \frac{ k_F m \Delta}{\pi \hbar^2 \sqrt{\omega^2-\Delta^2}}\\
\mbox{Im} \left[B^{(2)}(\omega)\right] &=&
-\frac{m^3 \Delta}{\hbar^6 \pi^3 R^2}
\left[ \frac{ k_F \omega}{\omega^2-\Delta^2} \sin^2{(k_F R)}
-\frac{1}{2 R_0 \sqrt{\omega^2-\Delta^2}} \sin{(2 k_F R)} \right]
.
\end{eqnarray}
Averaging over oscillations at scale $\lambda_F=2 \pi / k_F$ we obtain
the average value of $B^{(2)}(\omega)$:
\begin{equation}
\mbox{Im}\left[\overline{B^{(2)}(\omega)}\right]=
-\frac{m^3 \Delta k_F}{2 \hbar^6 \pi^3 R^2} \frac{\omega} {\omega^2-\Delta^2}
.
\end{equation}
The integral over $\omega$ of $g_{1,2}(\omega) + B^{(2)}(\omega) t^2$ is diverging
logarithmically and
the superconducting gap is reduced at order $t^2$. At order $t^4$ we obtain
the average value of $B^{(4)}(\omega)$:
\begin{equation}
\mbox{Im}\left[\overline{B^{(4)}(\omega)}\right]=
\frac{m^7 \Delta}{\hbar^{10} \pi^5 k_F R^2}
\left[ \frac{\omega^3 ( \omega+2 h_{\rm ex})}{(\omega^2-\Delta^2)^{3/2}}
-\frac{2 \omega h_{\rm ex}}{\sqrt{\omega^2-\Delta^2}}\right]
+ ...
\end{equation}
The perturbative
expansion of the self-consistent superconducting gap
at order $t^4$ breaks down
since $\overline{B^{(4)}(\omega)}$ grows like $\omega$. 
Therefore the integral over $\omega$ in (\ref{eq:self})
diverges faster than logarithmically.

\section{Bilayers and trilayers with atomic thickness}

Let us consider now infinite planar F/S interfaces~\cite{Buzdin-Daumens,Melin-Feinberg,Melin}.
For simplicity we restrict the
discussion to F/S bilayers and F/S/F trilayers with atomic thickness
(see Fig.~\ref{fig:schema}-(c)). We use the labels ``a'' and ``b'' for the two
ferromagnets and the label ``$\alpha$'' for the superconductor.
For strong ferromagnets ($\Delta_0 \ll t \ll h_{\rm ex}$)
the perturbative expansion to order $t_a^2$ and $t_b^2$ takes the form~\cite{Melin-Feinberg}
\begin{equation}
\ln{\left(\frac{\Delta}{\Delta_0}\right)}
=-2 \frac{t_a^2 + t_b^2}{h_{\rm ex}^2} \left[
\ln{\left(\frac{h_{\rm ex}}{\Delta_0}\right)}-\frac{1}{2} \right]
,
\end{equation}
where $\Delta_0$ is the superconducting gap of the isolated superconductor.
For weak ferromagnets ($t \ll h_{\rm ex} <\Delta_0$) we find~\cite{Melin-Feinberg}
\begin{equation}
\ln{\left(\frac{\Delta}{\Delta_0}\right)}
=-\frac{1}{2} \frac{ (t_a^2+t_b^2) h_{\rm ex}^2}{\Delta_0^4}
.
\end{equation}
At order $t^4$ the superconducting gap $\Delta_{\rm P}$
in the parallel alignment is different from the superconducting gap $\Delta_{\rm AP}$
in the antiparallel alignment. For strong ferromagnets we find~\cite{Melin-Feinberg}
\begin{equation}
\ln{\left(\frac{\Delta_{\rm P}}{\Delta_{\rm AP}}\right)}
=2 \frac{t_a^2 t_b^2}{h_{\rm ex}^4}
\left[ 7 \ln{
\left(\frac{h_{\rm ex}}{\Delta_0}\right)}
-4- \ln{\left(\frac{\Delta_0}{\eta}\right)} \right]
,
\end{equation}
where $\eta$ is an ultra-violet cut-off.
For weak ferromagnet we find~\cite{Melin-Feinberg}
\begin{equation}
\label{eq:expan-weak}
\ln{\left(\frac{\Delta_{\rm P}}{\Delta_{\rm AP}}\right)}
= 2 \frac{t_a^2 t_b^2}{\Delta_0^4}\left[
\frac{3}{2} + \ln{\left(\frac{4 h_{\rm ex} \eta}{\Delta_0^2}
\right)} \right]
+ 2 \frac{t_a^2 t_b^2 h_{\rm ex}^2}{\Delta_0^6}
\left[ -\frac{19}{6} + 2 \ln{\left(\frac{2 h_{\rm ex}^2}
{\Delta_0 \eta} \right)} \right] + ...
\end{equation}

\section{Conclusions}

We thus see that the perturbative expansions in the point contact
geometry are rather different from the trilayers with atomic thickness:
the small parameters are different, the divergencies are different. 
It was predicted for the point contact geometry that the self-consistent
superconducting gap is larger in the parallel alignment~\cite{Melin-FSF}.
This unusual feature
of the proximity effect is present also in the atomic thickness geometry
for half-metal ferromagnets~\cite{Buzdin-Daumens} and strong
ferromagnets~\cite{Melin-Feinberg}. However the case of weak ferromagnets 
cannot be reproduced with the point contact geometry.

For weak ferromagnets we obtained non monotonic temperature dependences of the
self-consistent superconducting gap for the F/S bilayer with atomic
thickness~\cite{Melin-Feinberg} and also with a finite thickness in the
ferromagnetic and superconducting electrodes~\cite{Melin}. This unusual
behavior is in agreement with the critical temperature obtained by
solving linearized Usadel equations~\cite{Baladie,Fominov}. This is related
to Andreev bound states in the middle of the superconducting
gap~\cite{Melin}. It also corresponds to the cross-over between the
perturbation theory of strong ferromagnets in the high temperature regime
(with $h_{\rm ex}>\Delta_0$)
and the perturbation theory of weak ferromagnets
(with $h_{\rm ex}<\Delta_0$) in the low temperature regime.

There are two differences between the point contact and atomic thickness
geometries: the localized versus extended nature of the contact and the lateral
confinement in the superconducting and ferromagnetic electrodes. 
A more complete investigation of the two factors will be presented elsewhere. We note 
here that from the
study of F/S/F trilayers with a finite thickness~\cite{Melin} we see that
the non monotonic temperature dependence of the self-consistent
superconducting gap exists if the width of the superconductor is smaller
than the Fermi wave-length $\lambda_F$ or larger than $\lambda_F$ but
smaller than the superconducting coherence length $\xi_0$. The width of the 
superconductor is thus a critical parameter for the appearance of 
the specific regime of weak ferromagnets. If the width is larger than
$\xi_0$ it can be conjectured that the the perturbation theory of the
infinite planar geometry on Fig.~\ref{fig:schema}-(b)
can be deduced from the summation of the perturbation theories of
a regular array of point contacts~\cite{Melin-FSF}. 
Finally geometrical effects were also investigated
in a recent work~\cite{Vecino} and the magnetization in the superconductor
was calculated recently~\cite{Volkov}, in agreement with the atomic thickness geometry
for strong ferromagnets. 
A more complete bibliography can be found elsewhere~\cite{Melin}.

\end{document}